\documentclass[aps,prb,twocolumn,amsmath,amssymb,superscriptaddress,floatfix,nofootinbib]{revtex4-2}

\usepackage{amssymb}
\usepackage{amsmath}
\usepackage{graphicx}
\usepackage{textcomp}
\usepackage{color}
\usepackage{xcolor}
\usepackage{comment}
\usepackage{amsfonts}
\usepackage{epsfig}
\usepackage{hyperref}
\usepackage{bm}
\usepackage[caption=false]{subfig} 
\usepackage{booktabs}
\usepackage{tabularx}
\usepackage{array}
\usepackage{soul}
\usepackage{wasysym}
\usepackage{feyn}
\usepackage{esint}
\usepackage{footmisc}
\usepackage{dcolumn}

\newcommand{\arctanh}[1]{\operatorname{arctan}}

\newcolumntype{M}[1]{>{\centering\arraybackslash}m{#1}}

\DeclareMathAlphabet\mathbfcal{OMS}{cmsy}{b}{n}

\begin{document}
%TC:ignore
\title{Spin-dependent transport in Fe${_3}$GaTe${_2}$ and Fe${_n}$GeTe${_2}$ ($n$=3-5) van der Waals ferromagnets for magnetic tunnel junctions}

\author{Anita Halder}
\email[]{haldera@tcd.ie}
\affiliation{School of Physics and CRANN, Trinity College, Dublin 2, Ireland}
\affiliation{Department of Physics and Centre for
Computational and Integrative Sciences, SRM University-AP, Amaravati 522 240, Andhra Pradesh, India}

\author{Declan Nell}
\affiliation{School of Physics and CRANN, Trinity College, Dublin 2, Ireland}

\author{Akash Bajaj}
\affiliation{School of Physics and CRANN, Trinity College, Dublin 2, Ireland}

\author{Stefano Sanvito}
\affiliation{School of Physics and CRANN, Trinity College, Dublin 2, Ireland}

\author{Andrea Droghetti}
\email[]{andrea.droghetti@unive.it}
%\affiliation{School of Physics and CRANN, Trinity College, Dublin 2, Ireland}
%\affiliation{Institute for Superconducting and other Innovative materials for devices, Italian National Research Council (CNR-SPIN), at G. d'Annunzio University, Chieti, Italy}
\affiliation{Department of Molecular Sciences and Nanosystems, Ca' Foscari University of Venice, via Torino 155, 30170, Mestre, Venice, Italy}

\begin{abstract}
We present a systematic first-principles investigation of linear-response spin-dependent 
quantum transport in the van der Waals ferromagnets Fe$_3$GeTe$_2$, Fe$_4$GeTe$_2$, 
Fe$_5$GeTe$_2$, and Fe$_3$GaTe$_2$. Using density functional theory combined with 
the non-equilibrium Green's function formalism, we compute their Fermi surfaces, transmission 
coefficients, and orbital-projected density of states. All compounds exhibit nearly half-metallic 
conductance along the out-of-plane direction. This is characterized by a finite transmission 
coefficient for one spin channel and a gap in the other, resulting in spin polarization values 
exceeding 90\% in the bulk. Notably, Fe$_3$GaTe$_2$ displays the ideal half-metallic behavior, 
with the Fermi energy located deep in the spin-down transmission gap. We further show that this 
high spin polarization is preserved in bilayer magnetic tunnel junctions, which exhibit a large 
tunnel magnetoresistance of the order of several hundred percent. This findings underscore the 
promise of these materials, and in particular of Fe$_3$GaTe$_2$, for spintronics applications. 
\end{abstract}

\maketitle

\section{Introduction}

Spintronics exploits the electron spin for faster, more efficient and nonvolatile electronic operations. 
A fundamental component of spintronics technology is the magnetic tunnel junction (MTJ), which 
consists of two ferromagnetic electrodes separated by a insulating barrier. MTJs implement the
tunnelling magnetoresistance (TMR) effect \cite{parkin, JULLIERE1975, MTJ-1, Yuasa2004}, wherein 
the electrical resistance of the junction changes upon reversing the relative alignment of the electrodes 
magnetization vectors. The TMR magnitude, and consequently the performance of MTJs, is primarily 
dictated by the spin polarization of the electrodes and their interfaces with the insulating 
barrier \cite{bu.zh.01, ma.um.01, PhysRevB.111.035133}. In principle, infinite TMR is achieved 
with half-metallic electrodes, where one spin channel is fully conductive and the other 
insulating \cite{Half-metal, coey-hm}. 

While conventional MTJs are fabricated using epitaxially grown thin films of bulk materials, such 
as the prototypical Fe/MgO system \cite{parkin, Yuasa2004}, the discovery of van der Waals (vdW) 
ferromagnets \cite{Gong2017, Huang2017, 2dmagnets1, 2dmagnets2} has opened new pathways 
for engineering fully vdW MTJs \cite{CrI3-MTJ1, CrI3-MTJ2}. In particular, compounds from the 
Fe$_n$GeTe$_2$ (F$n$GeT) ($n$ = 3, 4, 5) \cite{FGT41st-dup} family have recently emerged as 
the preferred electrode materials due to their robust metallic ferromagnetism, which persists down 
to the monolayer limit \cite{Fei2018, monolayer-FGT, FGT3-monolayer2}. Recent experiments have 
demonstrated a wide range of F$n$GeT-based MTJs using different 2D insulating barriers, including 
h-BN \cite{hBN,hbn2}, graphite \cite{graphite}, MoS$_2$ \cite{MoS2}, InSe \cite{InSe}, GaSe \cite{GaSe}, 
WSe$_2$ \cite{WSe2} and WS$_2$ \cite{WS}. These devices have shown TMR ratios up to 300\% at 
cryogenic temperatures, typically below 10 K \cite{hbn2,hBN}.

Among the F$n$GeT compounds, Fe$_3$GeTe$_2$ (F3GeT) was the first to be incorporated into an MTJ \cite{hbn2} and 
remains the most studied to date. It has been predicted to exhibit high spin polarization \cite{Tsymbol_mtjhbn} 
and displays relatively high perpendicular magnetic anisotropy \cite{FGT-mca}, which is essential for 
maintaining the magnetization alignment in the MTJs. However, the bulk Curie temperature, $T_\mathrm{C}$, 
of F3GeT is below room temperature, set at approximately 230 K \cite{FGT3, Fei2018}, and is further reduced 
in thinner layers, limiting its potential for applications. Increasing the Fe concentration, as in Fe$_4$GeTe$_2$ 
(F4GeT) and Fe$_5$GeTe$_2$ (F5GeT), raises the T$_\mathrm{C}$ to near or above room temperature 
\cite{FGT41st-dup,F5GT-R-3m,silinskas2024}. However, this enhancement comes at the cost of reduced perpendicular magnetic anisotropy.

These limitations of the F$n$GeT family appear to have been overcome in the recently reported vdW ferromagnet, 
Fe$_3$GaTe$_2$ (F3GaT), which is effectively the hole-doped version of F3GeT \cite{first_FeGaT_paper}. F3GaT 
is isostructural to F3GeT, but exhibits a remarkable $T_\mathrm{C}$ of 360 K, well above room temperature. Furthermore, 
it possesses an out-of-plane magnetic anisotropy significantly larger than that of its sister compound, F3GeT \cite{nanolett.aniso.mono.Ga}. 
F3GaT has been integrated into a full vdW F3GaT/WS$_2$/F3GaT MTJ yielding a high TMR of 213\% and a spin 
polarization of 72\% at 10 K. The TMR persists even at room temperature, although much lower, 11\% \cite{FGaT-WSe2exp}. 
Similar MTJs using WSe$_2$ as a barrier have achieved even higher TMR values of 340\% at 10 K and 
50\% at 300 K \cite{Parkin-et-al}.

On the theoretical front, numerous studies have explored the electronic and magnetic properties of F$n$GeT and 
F3GaT compounds \cite{Zh.Ke.16, Zh.Ja.16, Sh.Bo.21,  FGT41st-dup, sanyal-SR, FGT3-theory, F3GaT-mono-theory24, ErGh22}, 
while spin-dependent transport have been investigated for various model MTJs 
\cite{tsymbol-mtj, Tsymbol_mtjhbn, FGT-mtj-acs, FGT3-pra-mtj, FGT-mtj-mol, FGT4-bsanyal}. 
These studies largely agree that the F$n$GeT compounds exhibit high spin polarization for perpendicular-to-the-plane 
transport, with our previous study predicting that F4GeT may even approach half-metallic behavior \cite{our_nanoletter}. 
However, comparing these results from the literature is challenging due to variations in the considered system setups, 
material stacking, as well as in the computational details employed in the different studies. Consequently, it remains an 
open question which compound offers the highest spin polarization and is best suited for MTJ applications.

\begin{figure*}
\includegraphics[width=0.8\linewidth]{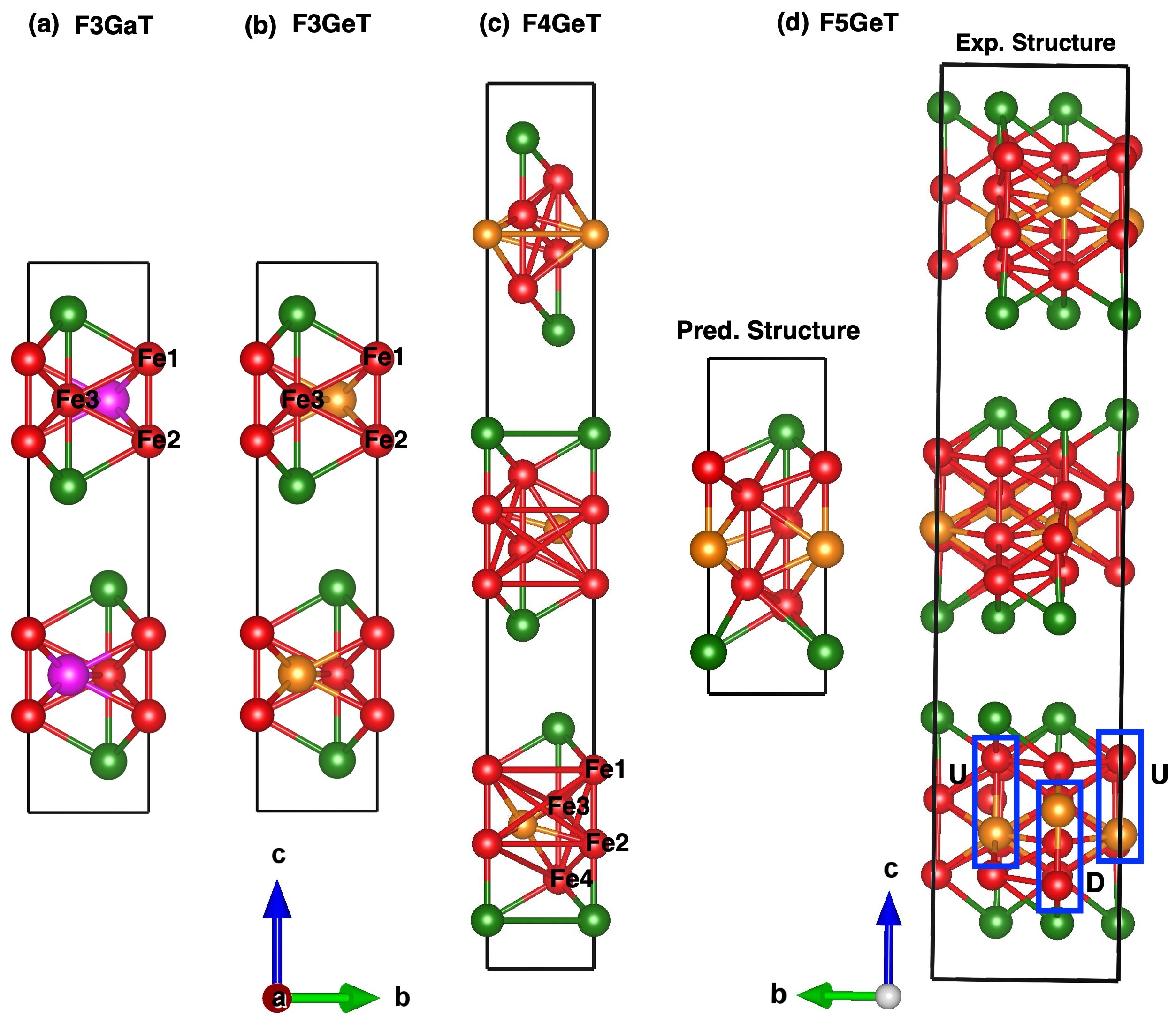}
\caption{Crystal structures of bulk (a) F3GaT, (b) F3GeT, (c) F4GeT, and (d) F5GeT as predicted theoretically (left) and 
observed experimentally (right). The Fe, Ge, Ga, and Te atoms are represented by red, orange, magenta, and green 
spheres, respectively. Inequivalent Fe atoms are explicitly indicated for F3GaT, F3GeT and F4GeT. The experimentally 
observed F5GeT structure is represented in the UUD configuration, with each Fe atom labeled as U or D depending 
on whether it is positioned above or below the Ge atom.}
\label{CS.comp}
\end{figure*}

In this work, we present a comprehensive and consistent comparison of the spin polarization of F3GeT, F4GeT, F5GeT 
and F3GaT using density functional theory (DFT) \cite{Kohn_nobel} combined with the non-equilibrium Green's function 
(NEGF) formalism for quantum transport \cite{Datta}. We evaluate the Fermi surface, transmission coefficients, and 
orbital-projected densities of states (DOS) for bulk systems. Our calculations reveal that all compounds exhibit nearly 
half-metallic conductance in the direction perpendicular to the layers, with spin polarization exceeding 0.9 and approaching 
unity in Fe$_3$GaTe$_2$. This high spin polarization is preserved in bilayer systems, which serve as model MTJs, where 
the two layers can be set in different magnetic states, and the vdW gap between them acts as insulating barrier. 
These bilayer MTJs exhibit TMR ratios of the order of several hundred percent, highlighting their strong potential for 
spintronics applications.

The paper is organized as follows. In Section \ref{sec.def}, we briefly review the DFT+NEGF method and introduce the 
definitions of conductance and spin polarization. Section \ref{sec.comput_details} provides the computational details. 
The results for bulk systems are presented next: we describe the crystal structures in Section \ref{sec.crystal}, analyze 
the Fermi surface and conduction channels in Section \ref{section: fermi_surface}, and examine the electronic structure 
of the different materials in Section \ref{sec.dos}. The bilayer systems used as MTJs are discussed in Section \ref{sec.bilayers}. 
Finally, we present our conclusions in Section \ref{sec.conclusion}.

\section{Spin-dependent quantum transport method}\label{sec.def}

Quantum transport calculations are performed using the DFT+NEGF method \cite{ro.ga.06, book1}. In particular, we 
consider transport perpendicular to the layers, along the Cartesian $z$ direction. The system is treated as infinite 
and it is partitioned into a central region coupled to two semi-infinite leads, described via self-energies \cite{ivan_self_energies.ss.08}. 
Periodic boundary conditions are applied in the $x$-$y$ plane, and $\mathbf{k}_\parallel = (k_x, k_y)$ denotes the Bloch 
wave vector in the two-dimensional (2D) Brillouin zone (BZ) in the transverse plane.

We adopt a two-spin-fluid picture \cite{Mott}, and perform spin-collinear calculations, following an established practice 
in the study of MTJs \cite{JULLIERE1975}. In this framework, coherent transport is determined by the transmission coefficient, 
$T^\sigma(E, \mathbf{k_\parallel})$, which describes the transmission of spin $\sigma=\uparrow\downarrow$ Bloch states 
having wave-vector $\mathbf{k}_\parallel$ from one lead, through the central region, into the other lead. The spin-dependent 
conductance in the linear-response limit at zero temperature is given by the Landauer-B\"uttiker formula \cite{La.57,Bu.86,Bu.88}
\begin{equation}
G^{\sigma} = G_0 T^{\sigma}(E_\mathrm{F})\:,
\end{equation}
where  
\begin{equation} \label{eqn: trc_total}
T^{\sigma}(E_\mathrm{F}) = \frac{1}{N_{\mathbf{k_\parallel}}} \sum_{\mathbf{k_\parallel}} T^\sigma(E_\mathrm{F}, \mathbf{k_\parallel})\:,
\end{equation}
is the total transmission coefficient at the Fermi energy, $E_\mathrm{F}$, with  ${N_{\mathbf{k_\parallel}}}$ being the 
total number of wave-vectors in the transverse BZ. Here, $G_0=\frac{e^2}{h}$ is the quantum of conductance, with 
$e$ is the electron charge and $h$ the Planck's constant. The transmission coefficient is calculated using the Fisher-Lee 
formula \cite{PhysRevB.23.6851}, as outlined, for example, in Refs. \cite{book1, dr.ra.22}. 

The linear response spin-polarization can be defined in terms of the spin-dependent conductance as
\begin{equation}\label{eqn: sp}
    \textnormal{SP} =\frac{G^{\uparrow} - G^{\downarrow}}{G^{\uparrow} + G^{\downarrow}}= \frac{T^{\uparrow}(E_\mathrm{F}) - T^{\downarrow}(E_\mathrm{F})}{T^{\uparrow}(E_\mathrm{F}) + T^{\downarrow}(E_\mathrm{F})}\:.
\end{equation}
This is the central quantity of interest for this paper, and it may range continuously from -1 to 1 (or from -100\% to 100\% in percentage terms). The two endpoints correspond 
to the half-metallic limit, where the conductance is finite for one spin channel but vanishes entirely for the other (by convention,
-1 corresponds to minority conductance, while +1 to majority).

\section{Computational details}\label{sec.comput_details}
\subsection{Geometry optimizations}
All geometry optimizations are performed using density functional theory (DFT) as implemented in the Vienna {\it Ab-initio} 
Simulation Package  ({\sc VASP})~\cite{VASP}. The Perdew-Burke-Ernzerhof (PBE) \cite{PBE} generalized gradient approximation 
(GGA) is adopted for the exchange-correlation functional, with vdW interactions treated at the DFT-D3 level \cite{DFTD3}. All 
calculations are spin-polarized. The atomic coordinates are relaxed until all atomic forces are smaller than 10$^{-3}$ eV/\AA. 
The Gaussian smearing method is employed, with a kinetic energy cutoff of 600~eV. A $\Gamma$-centered 
$10 \times 10 \times 1$ $\mathbf{k}$-mesh is used, along with a total energy convergence criterion of 10$^{-7}$ eV.

For bulk structures, only the atomic positions are relaxed, while the lattice parameters are maintained fixed to their 
experimental values. In the case of bilayers, a vacuum spacing of approximately 20~\AA~ was introduced along the 
$z$-direction to eliminate spurious interactions between periodic images. The optimal in-plane lattice parameters 
are determined through an energy-versus-volume calculation. After optimization of the atomic coordinates, the 
interlayer gap between of various different bilayer systems was found to vary between 2.91~\AA\ and 3.03~\AA.

\subsection{Quantum transport}
Quantum transport calculations are performed with the {\sc Smeagol} package \cite{ro.ga.06}, which combines 
the NEGF formalism with an in-house development version of the {\sc Siesta} DFT code \cite{siesta}. All DFT+NEGF 
calculations employed the PBE exchange-correlation functional \cite{PBE}. Core electrons are described using 
norm-conserving Troullier-Martins pseudopotentials \cite{Tr.Ma.91}, while the valence states are expanded in 
a multiple-$\zeta$ numerical atomic orbital basis set \cite{siesta}. 
The leads self-energy are calculated using the algorithm reported in Ref. \cite{ ivan_self_energies.ss.08}. An 
electronic temperature of 300 K is assumed, and the real-space integration grid corresponds to a plane-wave 
cutoff of 600 Ry.

The density matrix of the central region is calculated by integrating the lesser Green's function along a standard 
semicircular contour in the complex plane \cite{ro.ga.06}, using 16 points on both the semicircular arc and the straight 
segment along the imaginary energy axis, and including 16 Matsubara poles of the Fermi function. The density matrix 
is converged self-consistently on a $30 \times 30$ transverse $\mathbf{k}$-point grid. This is then used in a 
non-self-consistent calculation with a denser $100 \times 100$ $\mathbf{k}$-mesh to obtain the zero-bias DOS and 
transmission coefficient. All reported energies are shifted such that the Fermi energy is always set at 0~eV.

\section{Results for the bulk systems}

Quantum transport in bulk systems is modeled by replicating the materials' unit cell infinitely along the transport 
direction. In practice, this is achieved by embedding a single unit cell in the central region, connected to two semi-infinite 
leads composed of the same material, a setup that re-introduces translational symmetry along the transport direction. 
We now first detail the crystal structures of the materials used to construct our bulk systems, and then we provide a 
comprehensive analysis of their electronic structure and transport properties.
\begin{table}[h]
\centering
\begin{tabular}{|c|c|c|c|c|c|c|}
\hline
Compound  & Space group & a (\AA) & c (\AA) & Stacking  \\
\hline
F3GaT \cite{first_FeGaT_paper} & P63/mnc & 4.070 & 16.100 & A-B \\
\hline
F3GeT \cite{FGT5-intercalated} & P63/mnc & 3.990 & 16.300 & A-B  \\
\hline
F4GeT \cite{FGT5-intercalated} & R$\bar{3}$m & 4.040 & 29.080 & A-B-C \\
\hline
F5GeT (T)\cite{FGT5-intercalated} & P3m1 & 4.026 & 10.024 & A-A  \\
\hline
F5GeT (E)\cite{F5GT-R-3m, LyPa21} & P3$_2$ & 4.040 & 29.800 & A-B-C  \\
\hline
\end{tabular}
\caption{Crystal structure comparison of the vdW compounds considered in this work. 
F5GeT (T) and (E) refer to the theoretical and experimentally derived structures of F5GeT, 
respectively.}
\label{Crystal-structure}
\end{table}

\subsection{Crystal Structure}\label{sec.crystal}

The experimentally determined bulk crystal structures and in-plane and perpendicular lattice 
parameters, $a$ and $c$, are taken from Ref. \cite{first_FeGaT_paper} for F3GaT, and from 
Ref. \cite{FGT5-intercalated} for F3GeT and F4GeT, as summarized in Table \ref{Crystal-structure}. 
F3GaT and F3GeT are isostructural, with a lattice belonging to the P63/mnc space group with AB 
stacking \cite{F3GaT_F3GeT_isotructrutre}, while F4GeT is in the R$\bar{3}$m space 
group with ABC stacking \cite{FGT5-intercalated}. 

F3GeT and F3GaT contain three Fe atoms per layer within the unit cell, as illustrated in 
Fig.~\ref{Crystal-structure}, panels (a) and (b). Among these atoms, Fe1 and Fe2 are crystallographically 
equivalent and belong to the same sublattice, whereas Fe3 occupies a distinct inequivalent sublattice. 
The Ge ion is located at an interstitial site within the Fe sublattice, while the Te defines the top and 
bottom surfaces of each layer. Fe1 and Fe2 form a dumbbell-like structure oriented along the $c$-axis. 
In contrast, Fe3 lies in the same atomic plane as Ge and is vertically coordinated by the Te ions.

In comparison, F4GeT contains four Fe atoms per layer, occupying two crystallographically distinct 
sublattices, as shown in Fig. \ref{Crystal-structure}(c). Fe1 and Fe4 belong to one sublattice, while Fe2 
and Fe3 belong to the other, forming two pairs of Fe-Fe dumbbells. The Ge atom is located midway 
between Fe2 and Fe3, contributing to the in-plane coordination network. As in F3GeT, Te defines the 
top and bottom surfaces of each layer.

An ongoing ambiguity in the literature surrounds the structure of F5GeT. Initial experimental reports indicated that F5GeT crystallizes in the R3m space group \cite{FGT5-R3m}. Subsequent X-ray diffraction experiments, which probe the average structure, suggested instead that F5GeT belongs to the R$\bar{3}$m space group \cite{F5GT-R-3m, LyPa21, Wu.le.2021}. Further insight came from scanning tunneling microscopy (STM) studies \cite{F5GT-R-3m, LyPa21, Wu.le.2021, ZhNg23}, which revealed a $\sqrt{3} \times \sqrt{3}$R$30^\circ$ surface reconstruction \cite{F5GT-R-3m, LyPa21, Wu.le.2021}. 
Structurally, the reconstructed monolayer strongly resembles F4GeT, with the Ge atom residing at an 
interstitial site within the Fe sublattice, sandwiched between the Te layers. However, in F5GeT, an additional Fe atom is present along the axis normal to the vdW gap that passes through the Ge site.
In our coordinate frame, these Fe and Ge atoms form a subunit, which share the same in-plane coordinates 
($x, y$) but have different vertical, $z$, positions. The Fe atom may lie either above or below the Ge atom along the $z$-axis, positions that are referred to as the U (up) and D (down), respectively [see the right-side side of Fig. \ref{Crystal-structure}(d)]. Each $\sqrt{3} \times \sqrt{3}$R$30^\circ$ supercell contains three Fe/Ge subunits per layer, two in the U position and one in the D positions forming the so-called UUD configuration. This arrangement corresponds to the P$3_2$ space group, and has been shown in theoretical studies to be consistent with the STM observations \cite{ErGh22}.

However, previous quantum transport studies \cite{tsymbol-mtj} instead assumed a P3m1 space group with AA stacking \cite{FGT5-intercalated}, as shown on the left-hand side of Fig. \ref{Crystal-structure}(d). This monolayer is similar to the $\sqrt{3} \times \sqrt{3}$R$30^\circ$ reconstructed structure but with two main differences. First, all Fe/Ge subunits adopt the U configuration. Second, the bottom Te atoms are shifted in-plane so that their $(x,y)$ coordinates coincide with those of the Fe–Ge subunits.

In this work, for F5GeT, we consider both the P3m1 structure from previous quantum transport studies and 
the $\sqrt{3} \times \sqrt{3}$R$30^\circ$ reconstructed structure, which we refer to as the theoretical and 
experimentally derived structures, respectively. For the latter, we use experimentally determined lattice parameters 
and assume that bulk F5GeT adopts a ABC stacking sequence, with each layer in the UUD configuration. The 
resulting supercell contains 72 atoms, including 15 Fe atoms per layer.

\begin{figure*}[t]
\includegraphics[width=1.0\linewidth]{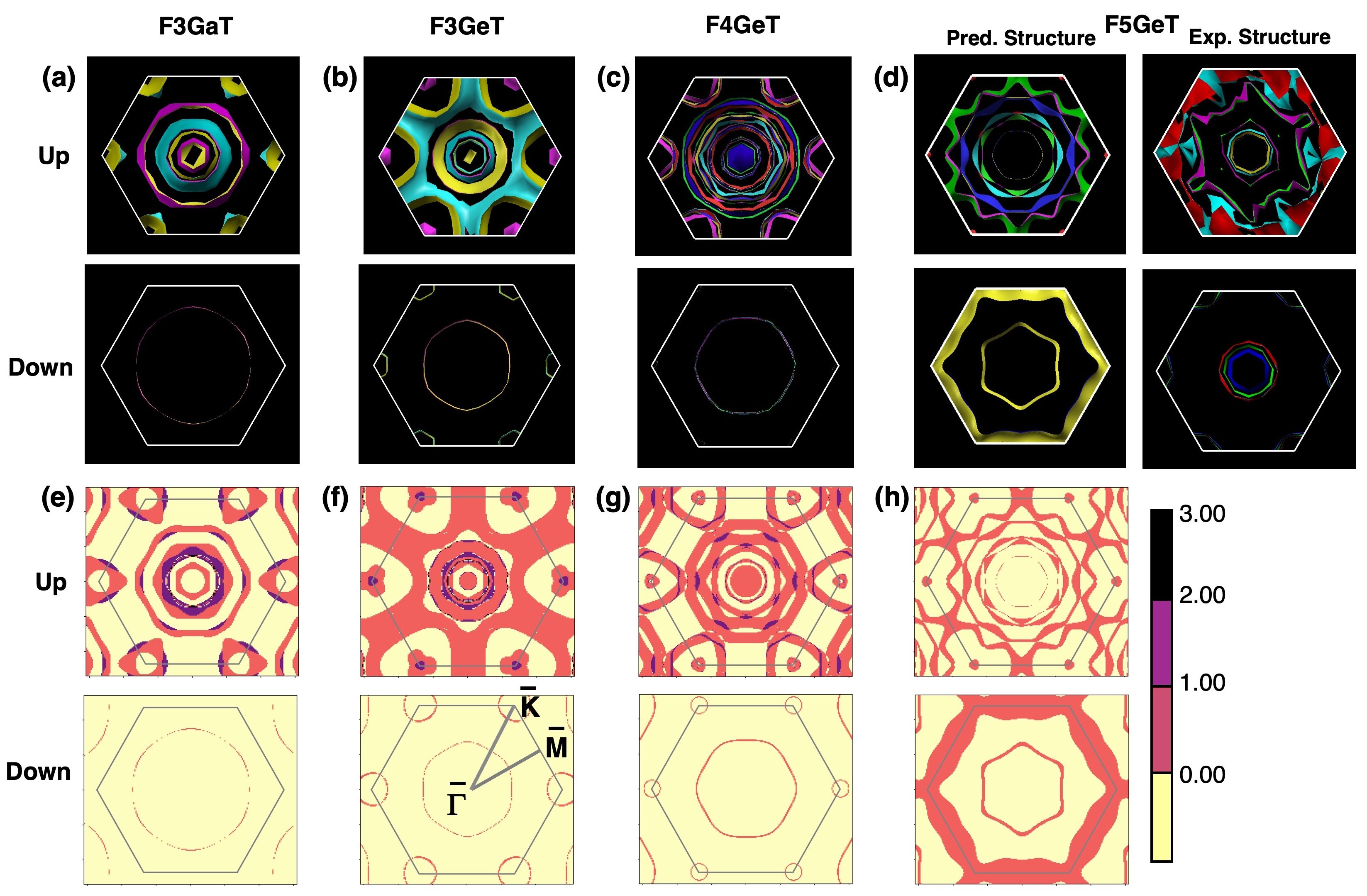}
\caption{Top six panels: Fermi surface for the spin up and spin down channels 
of (a) F3GaT, (b) F3GeT, (c) F4GeT and (d) F5GeT. Lower six panels: $\mathbf{k}_\parallel$-resolved 
zero-bias transmission coefficients at $E_{\mathrm{F}}$ for the spin-up and spin-down channels 
of (e) F3GaT, (f) F3GeT, (g) F4GeT and (h) F5GeT. The white hexagons indicates the BZ boundaries. }
\label{fermi_bulk.comp}
\end{figure*}

\subsection{Fermi surfaces and conduction channels} \label{section: fermi_surface}

%The key quantity for understanding spin-dependent transport is the number of conduction Bloch states (also referred to as ``channels'') available for each wave-vector $\mathbf{k_\parallel}$ in the 2D BZ. 

%In bulk systems, the leads and the central region are composed of the same material, resulting in ballistic transport with no electron scattering. The conducting Bloch states (also referred to as "channels") can see found from the material's Fermi surface projected onto the 2D BZ. Their number of conducting Bloch states (also referred to as "channels") for each transverse wave vector $\mathbf{k_\parallel}$ is given by the integer values of $T^\sigma(E_\mathrm{F},\mathbf{k_\parallel})$. 

\begin{figure}
\includegraphics[width=1.0\linewidth]{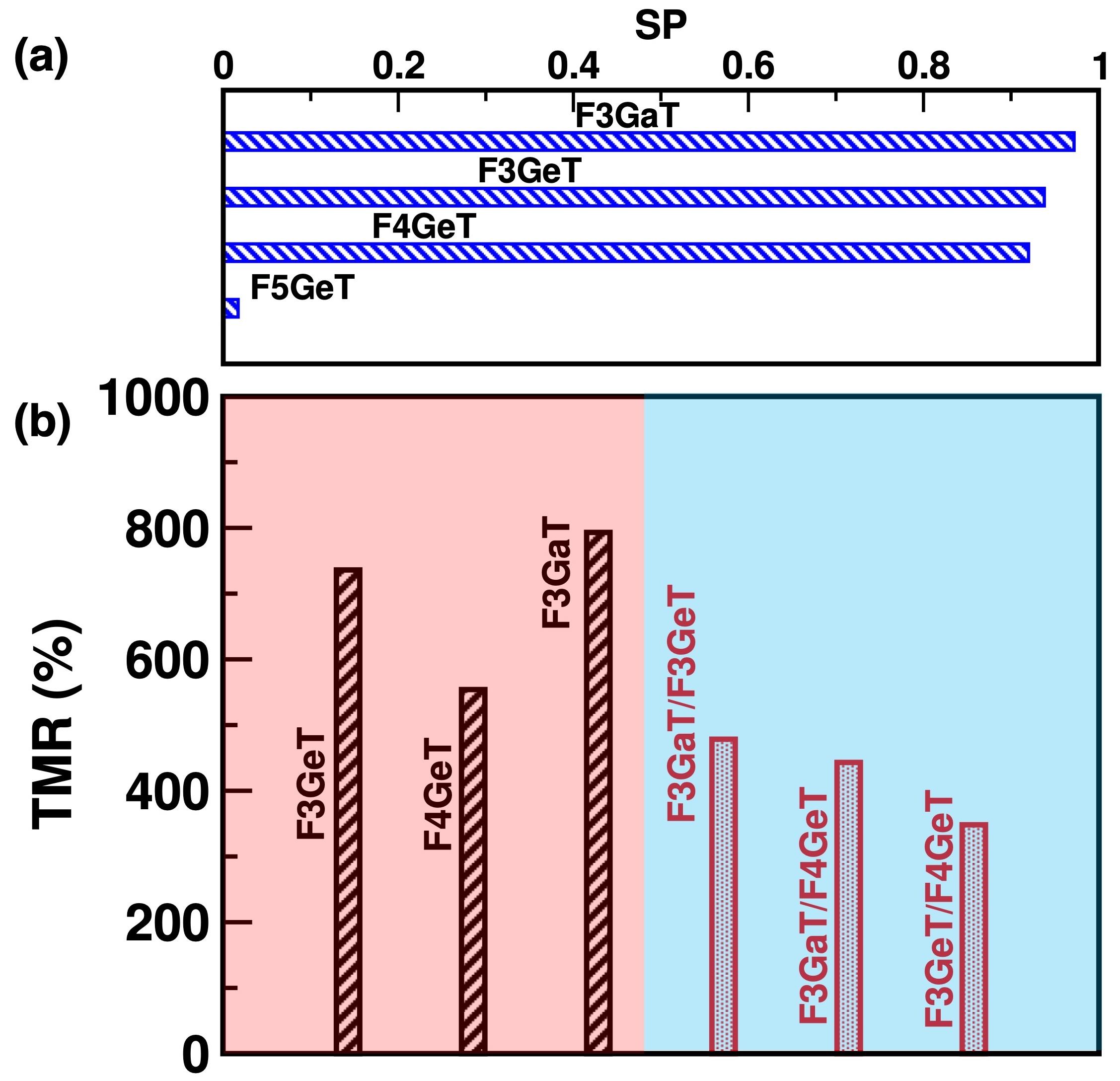}
\caption{(a) Bulk spin polarization (SP) for the investigated compounds. (b) TMR of bilayer MTJs formed by F3GaT, F3GeT, and F4GeT. Two types of MTJs are considered: (i) symmetric junctions, where both ferromagnetic layers are made of the same material (red shaded area), and (ii) asymmetric junctions, where the top and bottom layers are composed of different materials (blue shaded area).}
\label{TMR_SP}
\end{figure}

Electron transport is ballistic, namely scattering-free, in crystalline bulk systems, where both the leads and the central 
region are composed of the same material. The conducting Bloch states, also known as ``channels'', can be determined 
by plotting the material's Fermi surface projected onto the 2D BZ, as illustrated in Fig. \ref{fermi_bulk.comp} (top panels), 
where different colors correspond to distinct bands. For each transverse wave vector $\mathbf{k_\parallel}$, 
$T^\sigma(E_\mathrm{F},\mathbf{k_\parallel})$ corresponds to the the number of channels and therefore takes integer 
values. This is shown in Fig. \ref{fermi_bulk.comp} (bottom panels), where different colors indicate 0, 1, 2 or 3 channels. 
The total spin-up and spin-down transmission coefficients, and thus the SP, are then obtained by summing over all 
$\mathbf{k_\parallel}$, according to Eq. (\ref{eqn: trc_total}). In the following we analyze in details the results for each compound, 
with the calculated SP values summarized in Fig. \ref{TMR_SP}.

{\it F3GeT}. The spin-up Fermi surface spans a large portion of the 2D hexagonal BZ, as illustrated in the top panel 
of Fig. \ref{fermi_bulk.comp}(b). More specifically it features a central sheet at $\bar{\Gamma}$, surrounded by many 
concentric polygonal rings with an increasing number of sides as their radii increase. Furthermore, the radial sheets 
extend outward from the center of the BZ, reaching up the $\bar{\mathrm{M}}$ points, while distinct hexagonal sheets 
are observed around the $\bar{\mathrm{K}}$ points. In contrast, the spin-down Fermi surface is limited to a single 
ring around $\bar{\Gamma}$ and isolated hexagons around the $\bar{\mathrm{K}}$ points. These results are consistent with a previous report \cite{Tsymbol_mtjhbn}. 
Moving to $T^\sigma(E_\mathrm{F},\mathbf{k_\parallel})$, shown in Fig. \ref{fermi_bulk.comp} (f), it is clear that this
mirrors the Fermi surface and takes on a value of 1 in regions of the BZ where a single band crosses the Fermi level, 
and a value of 2 (occasionally 3) in regions where two (three) bands overlap. Thus, most conducting Bloch states at $E_\mathrm{F}$ are of 
spin-up character. This results in a SP as high as $94\%$, which approaches the half-metallic limit.

{\it F4GeT}. The spin-up Fermi surface [top panel in Fig. \ref{fermi_bulk.comp}(c)] exhibits a prominent circular feature 
at the $\bar{\Gamma}$ point, surrounded by concentric Fermi sheets forming polygonal rings. At the $\bar{\mathrm{K}}$ 
points, two bands overlap, resulting in a transmission coefficient of 2 at $\bar{\mathrm{K}}$. In contrast, the spin-down 
Fermi surface [bottom panel in Fig. \ref{fermi_bulk.comp}(c)] features only an isolated sheet around the $\bar{\Gamma}$ 
point, resembling that of F3GeT, and less visible sheets around the $\bar{\mathrm{K}}$ points. Consequently, as seen 
in Fig. \ref{fermi_bulk.comp}(g), most of the Bloch states conducting at $E_\mathrm{F}$ are of spin-up character, resulting 
in a SP of $92\%$. These results are consistent with those reported in our previous study on this compound \cite{our_nanoletter}.

%Notably, our calculations yield comparable SP values for F3GeT and F4GeT. This finding contrasts with the conclusions of a previous study \cite{our_nanoletter}, which suggested that F4GeT exhibited a higher spin polarization based on comparison calculations for F4GeT with data from the literature for F3GeT \cite{Tsymbol_mtjhbn} obtained using a different system setup and a computational implementation of DFT+NEGF. This demonstrates the critical importance of employing a consistent computational approach to reliably compare trends across different compounds.

%Several differences are notable. Firstly, F3GT exhibits an extended Fermi sheet along the $\bar{\Gamma}$-$\bar{\mathrm{M}}$ direction, which is instead absent in F4GT. Secondly, F3GT features a Fermi sheet covering each $\bar{\mathrm{K}}$ point, whereas F4GT displays a series of concentric polygonal rings around $\bar{\mathrm{K}}$. Lastly, the radius of the circle covering $\bar{\Gamma}$ is smaller for F3GT than for F4GT. Thus, in F4GT, the main contribution to spin-up transport originates primarily from channels near $\bar{\Gamma}$, unlike in F3GT, where there are also channels around $\bar{\mathrm{M}}$ and $\bar{\mathrm{K}}$.

{\it F5GeT.} The two structures considered -- namely the experimentally derived and the theoretically 
predicted one -- exhibit markedly different Fermi surfaces, as shown in Fig. \ref{fermi_bulk.comp}(d). 
For the experimental structure, the spin-up Fermi surface covers a large portion of the BZ, particularly 
near its boundaries. In contrast, the spin-down Fermi surface consists only of a few concentric rings 
centered around the $\bar{\Gamma}$ point. Although a full calculation of the transmission coefficient 
is not performed due to the large system size, the observed Fermi surface strongly suggests a very high 
SP, likely comparable to those of the other compounds. In contrast, the theoretically predicted structure 
displays a Fermi surface that spans substantial portions of the BZ for both spins. More in detail, the spin-up 
Fermi surface features several narrow rings with numerous wiggles centered around the $\bar{\Gamma}$ 
point, whereas the spin-down one comprises a single hexagonal ring centered at $\bar{\Gamma}$, accompanied 
by a broad sheet encircling the boundary of the BZ. Consequently, this theoretical structure exhibits a 
considerable number of open spin-down channels alongside the spin-up ones, despite their different distribution 
in the BZ, resulting in a negligible SP, of just about 2\%. %of $36\%$. This is considerably lower than that of the other compounds. Nevertheless, 
%we emphasize that this is not a particularly low SP value in absolute terms, as it remains comparable to SP 
%values reported for ferromagnetic metals such as Co in Co/Cu heterostructures \cite{dr.ra.22}. Thus, although 
%inferior to its related compounds, F5GeT in the theoretical structure still represents a reasonably good material 
%for spin transport.

{\it F3GaT.} The spin-up Fermi surface displays multiple sheets around $\bar{\Gamma}$, with one of the 
outer sheets taking a hexagonal shape aligned with the orientation of the BZ. In addition, triangular sheets, 
superimposed on the circular ones, are observed around the $\bar{\mathrm{K}}$ points. These features arise 
from the projection of the 3D Fermi surface at $\mathrm{K}$ and $\mathrm{A}$ in the 3D BZ onto the 2D one, 
as recently measured in angle-resolved photoemission spectroscopy experiments \cite{lee.ya.23}. In stark 
contrast, the spin-down Fermi surface [bottom panel of Fig. \ref{fermi_bulk.comp}(a)] consists of only a single 
large but narrow ring centered around $\bar{\Gamma}$, with an almost complete absence of states elsewhere.
In particular, unlike F3GeT, no spin-down bands are found in F3GaT around $\bar{\mathrm{K}}$. Consequently, 
the number of spin-down conducting channels is further reduced in F3GaT compared to the already small number 
in F3GeT. This reduction results in F3GaT achieving a SP as high as $97\%$. 

In summary, our calculations yield similarly high SP values for all compounds, except for the theoretical structure
of F5GeT. Notably, these results offer a different perspective than our previous study on F4GeT \cite{our_nanoletter}, 
whose SP was reported higher than that of F3GeT. However, that analysis relied on F3GeT literature data, obtained with different DFT+NEGF implementations and indirectly inferred from the results for systems presenting interfaces with various non-magnetic vdW materials \cite{Tsymbol_mtjhbn}.%, compared to those used in our calculations for F4GeT. 
This demonstrates the critical importance of employing a consistent computational 
approach to reliably compare trends across different compounds.

Among all studied systems, F3GaT has the highest SP, almost 100\%. This property, combined with the high 
$T_\mathrm{C}$ and out-of-plane magnetic anisotropy, makes F3GaT a truly unique vdW material for spintronics.

\begin{figure*}
\includegraphics[width=1.0\linewidth]{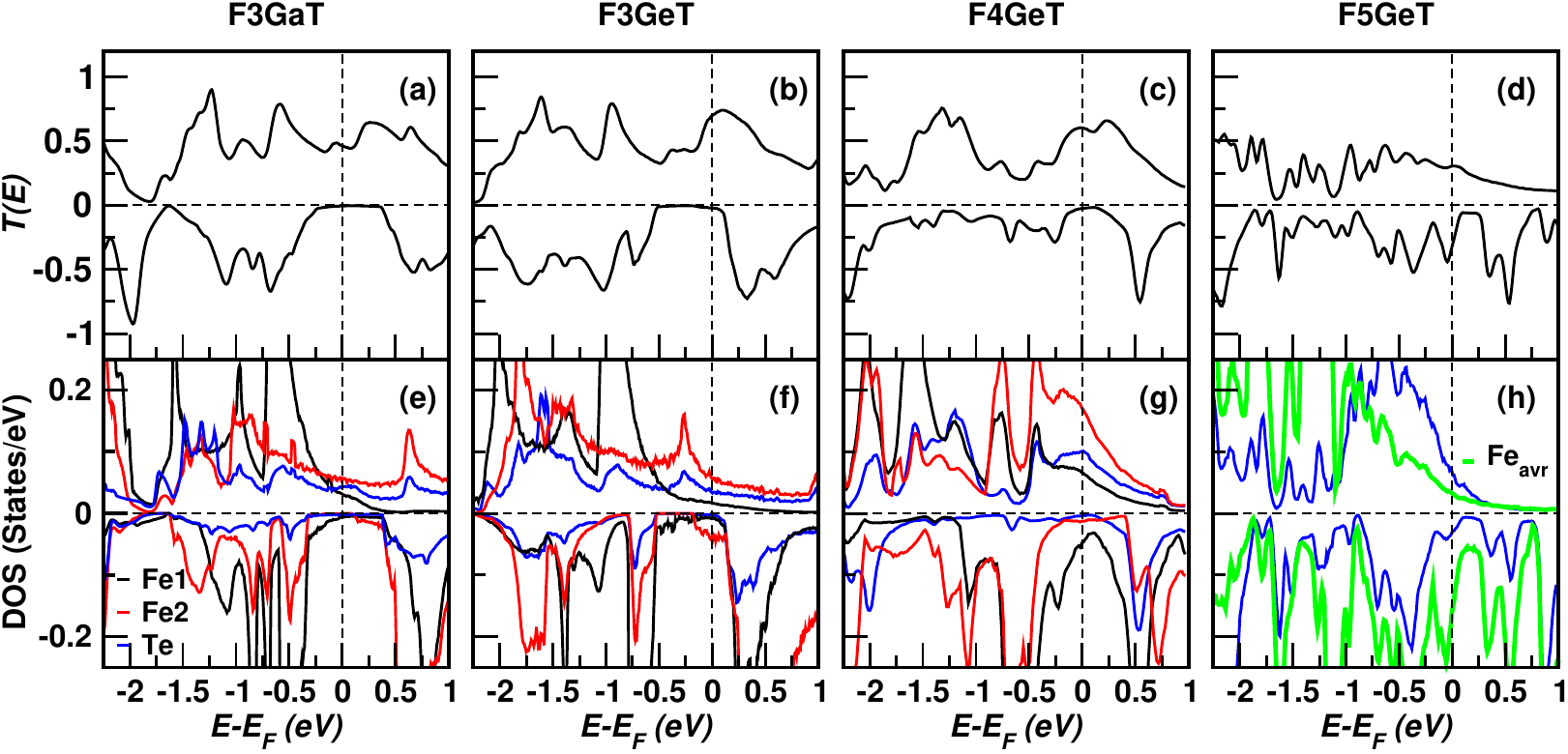}
\caption{Top panels: transmission coefficient at zero-bias for bulk 
(a) F3GaT, (b) F3GeT, (c) F4GeT, and (d) F5GeT. 
Bottom panels: PDOS for Fe1 $d_{z^2}$ (black), Fe3 $d_{z^2}$ (red), the average over all Fe $d_{z^2}$ orbitals (green), and Te $p_{z}$ (blue) in bulk (e) F3GaT, (f) F3GeT, (g) F4GeT, and (h) F5GeT. The labeling of inequivalent Fe atoms follows the convention introduced in Fig.~\ref{CS.comp}. Spin-up (down) values are shown positive (negative).}
\label{TRC_dos_fgt}
\end{figure*}

\subsection{DOS and transmission coefficients} \label{sec.dos}

The calculation of the zero-bias transmission coefficients can be extended beyond $E_\mathrm{F}$ over 
a broad energy range and compared with the DOS, providing a more complete view of the electronic structure of the materials. 
This comparison is shown in Fig. \ref{TRC_dos_fgt}, where the DOS is projected onto the out-of-plane 
Te $5p_z$ orbitals and the $3d_{z^2}$ orbitals of the inequivalent Fe atoms. As detailed in Sec. \ref{sec.crystal}, 
Fe1 and Fe2 are equivalent while Fe3 is distinct in F3GaT and F3GeT. For F4GeT, the equivalent pairs are 
(Fe1, Fe4) and (Fe2, Fe3). For F5GeT, the experimental structure is omitted due to its complexity, whereas in the theoretical structure all Fe atoms are inequivalent, and we therefore present the averaged (avr) Fe PDOS.

In F3GeT/GaT and F4GeT, $T^{\uparrow}(E)$ exhibits a prominent peak around $E_\mathrm{F}$, while 
$T^{\downarrow}(E)$ shows a gap in the same region. This shape is characteristic of nearly half-metallic 
transport, consistent with the Fermi surface analysis. The transmission coefficients closely follow the 
Te $5p_z$-PDOS, particularly in the spin-down channel, indicating that perpendicular transport primarily 
occurs through these orbitals, which are strongly hybridized with the $d_{z^2}$ of Fe3 and thus spin-split. 
Since Te ions are at surfaces of each layer, their inter-layer overlap determines the effective delocalization 
of the conducting electronic states.

In F3GeT, the spin-down gap arises from the splitting between the Te $p_{z}$ and Fe $d_{z^2}$ bonding and 
antibonding states, which is nearly 1~eV. However, this gap is effectively reduced to approximately 0.6~eV due 
to the electronic broadening. A similar behavior is observed in F3GaT, which can be regarded as a nearly hole-doped 
counterpart of F3GeT, as evident from a comparison of the PDOS of the two materials [Figs. \ref{TRC_dos_fgt}(e) and 
\ref{TRC_dos_fgt}(f)]. The transmission coefficient of F3GaT can be approximated by that of F3GeT, with the 
spin down gap shifting from $-0.55\;\textnormal{eV} < E-E_{\mathrm{F}} < 0.1\;\textnormal{eV}$ in F3GeT to 
$-0.25\;\textnormal{eV} < E-E_{\mathrm{F}} < 0.4\;\textnormal{eV}$ in F3GaT, as seen in Figs. \ref{TRC_dos_fgt}(a) 
and \ref{TRC_dos_fgt}(b). Importantly, $E_\mathrm{F}$ lies deep within the spin-down gap of F3GaT, whereas it 
is only 0.1~eV below the conduction states of F3GeT. It is this deeper $E_\mathrm{F}$ position which stabilizes 
the half-metallic character of F3GaT, making it an almost ideal material for spin transport. 

In F4GeT, the spin-down gap in the transmission coefficient [Fig. \ref{TRC_dos_fgt}(c)] is less sharply defined 
when compared to that of F3GaT and F3GeT. This is mostly due to large electronic broadening of the $d_{z^2}$ Fe1 
states, as seen in their PDOS [Fig. \ref{TRC_dos_fgt}(g)], induced by the bond with the equivalent Fe4 atom, absent in 
the other compounds. The results are in agreement with previous studies \cite{our_nanoletter}. Although the SP 
at $E_\mathrm{F}$ in F4GeT is comparable to that of F3GeT, the less well-defined spin-down gap makes its 
half-metallic behavior in principle more susceptible to degradation. Nevertheless, calculations that include various 
potentially detrimental effects, such as spin-orbit coupling, electron correlation, and homogeneous disorder, indicate 
that the system's high SP remains large in practice \cite{our_nanoletter}.

Finally, F5GeT (theoretical structure) also exhibits a spin-down gap in $T^\downarrow(E)$ [Fig. \ref{TRC_dos_fgt}(d)], similar to the other compounds. However, this gap is significantly reduced in size and shifted upward in energy, extending from $E-E_{\mathrm{F}} \approx 0.1$ eV to $E-E_{\mathrm{F}} \approx 0.25$ eV. As a result, $E_\mathrm{F}$ lies just below the gap edge, within states of predominantly Fe character [Fig. \ref{TRC_dos_fgt}(h)], leading to the negligible spin polarization predicted by the Fermi-surface analysis.

%In summary, the analysis of the energy dependence of the PDOS and transmission coefficients confirms that F3GeT, F4GeT, and F3GaT exhibit comparable spin transport properties and overall similar electronic structures. However, F3GeT stands out as the material closest to ideal half-metallic transport behavior, with the Fermi energy positioned deep within the spin-down gap in the transmission coefficient.
%Notably, the fact that the electronic structure in similar across across such a wide range of stoichiometries and geometries also suggests that nearly half-metallic transport in this family of compounds is likely robust against defects and disorder, and .

In summary, the analysis of the energy dependence of the PDOS and transmission coefficients confirms that F3GeT, 
F4GeT, and F3GaT exhibit overall similar electronic structures, characterized by a sizable gap in the spin-down channel. 
However, the position of the Fermi energy relative to the spin-resolved transmission gap can be further optimized through 
stoichiometric tuning. In this regard, F3GeT already stands out as the material closest to ideal half-metallic behavior, 
with its $E_\mathrm{F}$ lying deep within the spin-down gap

\begin{figure}
\includegraphics[width=1.0\linewidth]{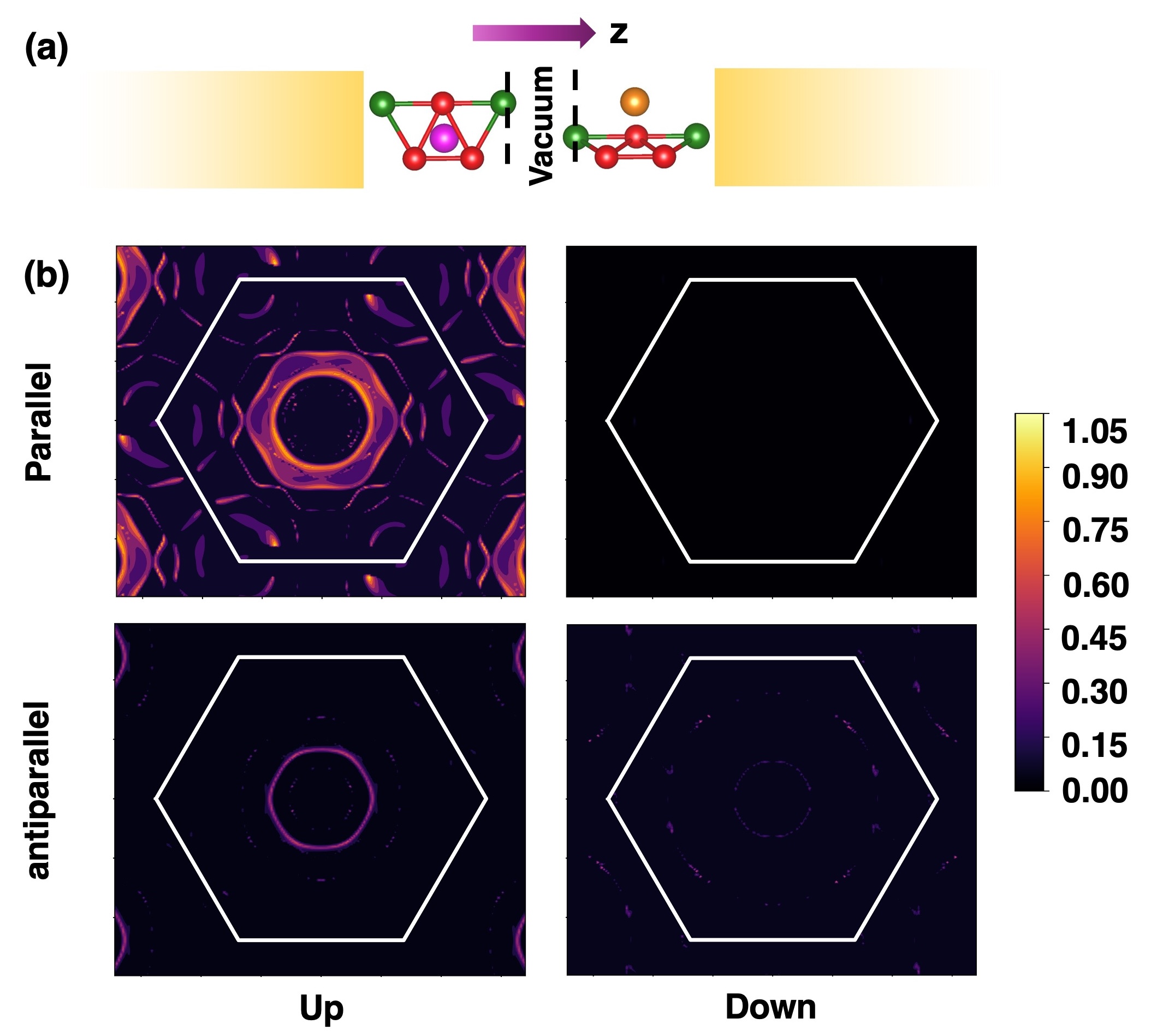}
\caption{F3GaT/F3GeT bilayer MTJ. (a) Schematic structure of the junction, with model leads represented as 
semi-infinite yellow bars. (b) Spin- and $\mathbf{k}_{\parallel}$-resolved transmission coefficients at $E_\mathrm{F}$ 
for the P and AP states. The white hexagons indicates the BZ boundaries. }
\label{F3GaT.mtj}
\end{figure}

\section{Results for bilayer MTJ\lowercase{s}}\label{sec.bilayers}

Model MTJs can be realized by connecting two F$n$Ge/GaT layers to generic metallic leads, with the vdW 
gap serving as the insulating barrier, as shown in Fig.~\ref{F3GaT.mtj}(a). 
The generic metallic leads are computationally simulated as a lattice of Au atoms using only the $6s$ orbitals as a basis, analogous to Ref.~\cite{our_nanoletter}. 
The bilayer can be in either 
the parallel (P) or antiparallel (AP) magnetic state, determined by the relative orientation of the magnetization 
vectors in the ferromagnetic layers. The two ferromagnetic layers may consist of either the same compound 
or different combinations of compounds. The latter case is particularly relevant for potential experimental 
realizations, as different layers will have different coercive fields, enabling independent switching of their 
magnetization directions. Accordingly, we consider homo-bilayers of F3GaT, F3GeT, and F4GT as well 
as the bilayer heterostructures F3GaT/F3GeT, F3GaT/F4GeT, and F3GeT/F4GeT. 

The linear-response TMR ratio is defined as
\begin{equation}\label{eq: tmr}
    \textnormal{TMR} = \frac{T_{\textnormal{P}}(E_\mathrm{F})-T_{\textnormal{AP}}(E_\mathrm{F})}{T_{\textnormal{AP}}(E_\mathrm{F})},
\end{equation}
where $T_{\textnormal{P}}$ ($T_{\textnormal{AP}}$) is the transmission coefficient for the P (AP) configuration. This 
corresponds to the so-called optimistic definition, which assumes that $T_{\textnormal{AP}}<T_{\textnormal{P}}$.

The calculated TMR values for all systems are shown in Fig. \ref{TMR_SP}(b). The highest TMR is obtained 
for the F3GaT bilayer MTJ, reaching up to nearly 800\%. F3GeT exhibits a slightly lower value of about 750\%, 
consistent with its slightly smaller spin polarization. In contrast, F4GeT shows a significantly reduced TMR of 
about 450\%, in agreement with the predictions of our previous work \cite{our_nanoletter}.

Compared to the MTJs composed of identical layers, the bilayer heterostructures exhibit a somewhat lower TMR. 
However, the calculated values remain quite high, around 400\% in all cases, with the F3GaT/F3GeT combinaiton 
reaching up to nearly 500\%. This is due to the fact that the electronic structures of the two materials are overall 
quite similar at the Fermi level.

The spin and $\mathbf{k}_{\parallel}$-resolved transmission coefficients at $E_{\mathrm{F}}$ of the F3GaT/F3GeT 
MTJ in both the P and AP state is shown in Fig. \ref{F3GaT.mtj}(b). Unlike the ballistic transport in bulk systems, 
discussed in Sec. \ref{section: fermi_surface}, the transmission coefficients here exhibit non-integer values due to 
elastic scattering occurring at the interfaces between the model leads and the bilayer, as well as between the two 
different ferromagnetic layers. In the P state, $T^\uparrow_\mathrm{P}(E_\mathrm{F}, \mathbf{k}_{\parallel})$ is 
primarily concentrated within a circle surrounded by a pentagon around the $\Gamma$ point, where it can can 
reach values of about 0.8. In contrast, $T^\downarrow_\mathrm{P}(E_\mathrm{F}, \mathbf{k}_{\parallel})$ is nearly 
zero across the entire BZ, reflecting the presence of a gap in the PDOS for the spin-down band of both layers, 
similar to that shown in Fig. \ref{TRC_dos_fgt}. 

In the AP state, the transmission also originates predominantly from the spin-up channel (this is defined with respect
to the first layer). However, $T^\uparrow_\mathrm{AP}(E_\mathrm{F}, \mathbf{k}{\parallel})$ exhibits a single circular feature in the 
BZ, with values reduced to around 0.3. To a first approximation, this $T^\uparrow_\mathrm{AP}(E_\mathrm{F}, \mathbf{k}{\parallel})$
can be approximated as the convolution of the spin-up transmission of F3GaT with the spin-down transmission of F3GeT
(see Fig.~\ref{fermi_bulk.comp}). 
In turn, $T^\downarrow_\mathrm{AP}(E_\mathrm{F}, \mathbf{k}_{\parallel})$ can be understood as the convolution of the 
spin-down transmission of F3GaT with the spin-up transmission of F3GeT. Since for both compounds the spin-down 
transmission is negligible, the total AP transmission $T_\mathrm{AP}(E_\mathrm{F}, \mathbf{k}_{\parallel})$ remains 
low, resulting in a large TMR.

In summary, the transport results confirm the potential of the F$n$Ge/GaT family for use in MTJs, demonstrating 
that even experimentally relevant heterojunctions can exhibit TMR values on the order of several hundred percent, 
driven by the peculiar electronic structure of these materials.

\section{Conclusion}\label{sec.conclusion}

We systematically investigated the spin-dependent transport properties of the vdW ferromagnets F3GeT, 
F4GeT, F5GeT and F3GaT by calculating their Fermi surfaces, transmission coefficients, and PDOS using 
DFT+NEGF.

Our results show that all these materials, with the exception of F5GeT in the theoretical structure, exhibit nearly 
half-metallic conductance in their bulk form, with spin polarization values exceeding 0.9 for transport perpendicular 
to the layers. This owns to the appearance of a sizebale gap in the spin-down PDOS. Notably, the fact that high 
SP is preserved across such a wide range of stoichiometries and geometries also suggests that the nearly half-metallic 
transport in this family of compounds is likely to be quite robust against defects and disorder.

Among the compounds studied, F3GaT emerges as the most promising candidate, with a SP approaching unity 
due to the Fermi energy lying deep within the spin-down transmission gap. This property, combined with its remarkable 
T$_\mathrm{C}$ and out-of-plane magnetic anisotropy reported in the literature \cite{first_FeGaT_paper,nanolett.aniso.mono.Ga}, 
makes F3GaT an exceptional vdW material for spintronics.

The high SP of F3Ge3T, F3GaT and F4GeT is retained also for bilayer systems, used as model MTJs, resulting in large 
TMR values. In particular, the homo-bilayer F3GaT is predicted to achieve a TMR ratio of about 900\%. Moreover, TMR 
values of several hundred percent are retained even in experimentally relevant heterojunctions formed by stacking layers 
of different materials, whose magnetization vectors can, in principle, be switched independently of each other. 
These findings are expected to guide further experimental and theoretical research toward the development of vdW-based 
MTJs.

\section{Acknowledgment}
A.H. was supported by European Commission through the Marie Sk\l{}odowska-Curie individual fellowship VOLTEMAG-101065605 and from the ANRF-PMECRG under sanction order ANRF/ECRG/2024/001931/PMS. D.N. was supported by the Irish Research Council (Grant No. GOIPG/2021/1468). 
A.B. and S.S. were supported by SFI (19/EPSRC/3605) and by the Engineering and Physical Sciences Research Council (EP/S030263/1).
Computational resources were provided by Trinity College Dublin Research IT and the Irish Center for High-End Computing (ICHEC).

\bibliography{main}

\end{document}